\documentclass[aps,showpacs,superscriptaddress,groupedaddress]{revtex4}  
\usepackage{graphicx}  
\usepackage{dcolumn}   
\usepackage{bm}        
\usepackage{amssymb}   
\usepackage{amsmath}
\usepackage{lipsum} 
\usepackage{braket}
\usepackage{siunitx}
\usepackage{booktabs} 
\usepackage{multirow}
\usepackage{hyperref}
\hypersetup{colorlinks  = true,
	urlcolor    = blue, 
	linkcolor   = blue, 
	citecolor   = red 
}
\usepackage{cleveref}

\begin{document}
	
	\title{Depolarization of Spin-Polarized hydrogen via Collisions with Chlorine Atoms at Ultrahigh Density}
	
	\author{Alexandros K. Spiliotis}
	\author{Michalis Xygkis}
	\affiliation{Foundation for Research and Technology Hellas, Institute of Electronic Structure and Laser, N. Plastira 100, Heraklion, Crete, Greece, GR-71110}
	\affiliation{University of Crete, Department of Physics, Herakleio, Greece}
	\author{Michail E. Koutrakis}
	\affiliation{University of Crete, Department of Physics, Herakleio, Greece}
	\author{Dimitrios Sofikitis}
	\affiliation{University of Ioannina, Department of Physics, Ioannina, Greece}
	\author{T. Peter Rakitzis}
	\email{ptr@iesl.forth.gr}
	\noaffiliation	
	\affiliation{Foundation for Research and Technology Hellas, Institute of Electronic Structure and Laser, N. Plastira 100, Heraklion, Crete, Greece, GR-71110}
	\affiliation{University of Crete, Department of Physics, Herakleio, Greece}
	
	\begin{abstract}
		Recently, the production of ultrahigh-density ($\sim10^{19} cm^{-3}$) spin-polarized deuterium (SPD) atoms was demonstrated, from the photodissociation of deuterium iodide, but the upper density limit was not determined. Here, we present studies of spin-polarized hydrogen (SPH) densities up to $10^{20}$\si{\per\centi\meter\cubed}, by photodissociating 5 bar of hydrogen chloride with a focused 213 nm, 150 ps laser pulse. We extract the depolarization cross-section of hydrogen and chlorine atom collisions, which is the main depolarization mechanism at this high-density regime, to be  $\sigma_{H-Cl}=7(2)\times10^{-17}$\si{\centi\meter\squared}. We discuss the conditions under which the ultrahigh SPH and SPD densities can be reached, and the potential applications to ultrafast magnetometry, laser-ion acceleration, and tests of polarized nuclear fusion.
	\end{abstract}
	
	\maketitle
	
	\section{Introduction}
	
	Polarized atomic gases typically have low densities, of $\sim10^{12}$\si{\per\centi\meter\cubed} and below, because of depolarizing effects of collisions in the production methods of Stern-Gerlach spin separation \cite{Szczerba_2000}, or because of optical-thickness
	effects in optical pumping \cite{Clasie_2006, SEOP_SPH}, due to the large absorption cross sections ($\gtrsim 10^{-10}$\si{\centi\meter\cubed}). The only exceptions have been the noble gases with nuclear spin (e.g. ${}^3He$ and ${}^{129}Xe$), for which the nuclear spins can be polarized at high pressure (via spin-exchange optical pumping), as the depolarization rate of these closed-shell atoms can be very small \cite{Walker_1997}. 	However, this inability to produce high-density open-shell spin-polarized gases precludes several potential applications, which are discussed below.
	
	Recently, Sofikitis \textit{et al.} demonstrated the production of ultrahigh-density ($\sim 10^{19}$ $cm^{-3}$), highly spin-polarized hydrogen (SPH) and deuterium (SPD), from the photodissociation of hydrogen halides with a focused, circularly polarized UV laser pulse, using a pickup coil to monitor the electron polarization \cite{Sofikitis_HighDensity}. These densities surpass the current	state-of-the-art by at least 7 orders of magnitude. To achieve such high densities, the hydrogen halides were photodissociated with a 150 ps laser pulse, so that the SPH/SPD are produced nearly instantaneously, on the $10^{-10}$s timescale, several orders of magnitude faster than the production time for conventional polarization methods. This rapid production allows the high-density SPH/SPD to be observed with 5-100 ns lifetimes at these high densities \cite{Boulogiannis_2019}, before depolarization from collisions. In addition, the HCl photodissociation cross section at 213 nm is $\sim10^{-21}$ \si{\centi\meter\squared}, which is small enough to avoid significant optical-thickness effects for densities as high as $10^{20}$ \si{\centi\meter\cubed}, but large enough to allow complete photodissociation of the HCl molecules for a focused UV beam of a few mJ. This is also an important factor for why SPH production from hydrogen-halide photodissociation can surpass traditional production methods by many orders of magnitude.
	
	The production of $\sim10^{15}$ SPH per laser shot, within the detection volume and on the sub-ns timescale, allows improvement in time resolution of atomic magnetometers by several orders of magnitude, and with spatial resolution of about 10 \si{\micro\meter} \cite{Spiliotis_Magnetometry}.	Laser-ion acceleration can produce intense, GeV-scale, electron or proton pulses, from gas targets of density $\sim10^{19}$\si{\per\centi\meter\cubed}, which have been unpolarized; only SPH production of HY photodissociation has demonstrated polarized electron, proton, and deuteron densities of $\sim$$10^{19}$\si{\per\centi\meter\cubed}, sufficient for laser-ion acceleration \cite{Wu_2019, Wen_2019, Jin_2020, Buscher_2020}. It is known that polarized nuclei increase the fusion cross sections of the D+T and D + $^{3}He$ reactions by $\sim$50\%, whereas the effect for the D + D reaction is not known. DI photodissociation can offer sufficiently dense SPD for tests of polarized laser fusion for these three reactions \cite{Sofikitis_SPD_PRL,Sofikitis_HighDensity, Spiliotis_LSA}. The signal for the D + D reaction is quadratic in the SPD density (whereas the dependence is linear for the other two reactions), therefore the SPH/SPD density limits determined in this paper are important for calculation of potential reaction yields.
	
	The aim of this paper is to investigate the main depolarization mechanisms in detail, and to determine the limits of SPH density and lifetime, to help understand the limits of applicability of SPH described above. In section \ref{sec:mechanism}, we discuss the main SPH depolarization mechanisms at ultrahigh densities of SPH produced via photodissociation of hydrogen halides. Section \ref{sec:experiment} describes the experimental procedure, as well as the method used to extract the depolarization cross-section. Finally, in section \ref{sec:results}, a value of the depolarization cross-section is determined, complemented with a discussion on the effect of a buffer gas(here $C_2F_6$) on the depolarization rate, and finally, a discussion on the expected polarization degree of protons and deuterons produced via the photodissociation method.
	  	
	\section{Depolarization mechanism}\label{sec:mechanism}
	The work presented in Sofikitis \textit{et al.} \cite{Sofikitis_HighDensity} and Boulogiannis \textit{et al.} \cite{Boulogiannis_2019} demonstrate two clearly distinct photodissociation regimes, depending on the photon density of the photodissociating pulse. At the "low" photon density, produced using the unfocused laser beam, only $\approx$0.1\% hydrogen halide molecules are dissociated to produce SPH and halogen atoms, which are then surrounded by HY atoms. We note, however, that this "low" SPH density of $\sim10^{16}$ \si{\per\centi\meter\cubed} is still a few orders of magnitude higher than SPH densities produced by conventional production methods. When the photon density is high enough that virtually all hydrogen halide molecules are dissociated, the  photodissociation volume is occupied solely by the molecular fragments, namely SPH and halogen atoms, which then interact with each other. This "high" photon density can be readily achieved for the same pulse by using a suitable lens to tightly focus the pulse inside the pickup coil. Assuming hydrohalide pressure of order bar, and a pulse energy of several mJ, the "low" SPH densities are produced with a beam with a diameter of order \si{\milli\meter}, whereas the "high" SPH density can be achieved when the pulse is focused to a diameter of $\sim$10 \si{\micro\meter}, e.g. by a f=50 \si{\milli\meter} lens.
	
	Boulogiannis \textit{et al.} showed that, in the low photon density limit, the main depolarization mechanism of SPH is via the formation of a short-lived SPH-HY complex, followed by intramolecular depolarization via the spin-orbit interaction: 
	
	\begin{align}
		& \label{complex_formation} SPH + HY \rightarrow HY\mspace{-8.0mu}-\mspace{-7.0mu}SPH\\ 
		& \label{complex_depolarization} HY\mspace{-8.0mu}-\mspace{-7.0mu}SPH \rightarrow HY\mspace{-8.0mu}-\mspace{-8.0mu}H \\
		& \label{complex_dissociation} 
		HY\mspace{-8.0mu}-\mspace{-7.0mu}SPH + HY \rightarrow SPH + 2HY
	\end{align}
	where HY-SPH and HY-H denote a complex formed by a hydrogen halide and a spin -polarized or unpolarized hydrogen atom, respectively.
	
	In the high photon density limit, which provides the upper limit of the SPH density for a given pressure, SPH depolarization results from collisions between the hydrogen and halide fragments:
	
	\begin{align}
		& SPH + Y \rightarrow \textrm{(relaxation channel)} \rightarrow H + Y\label{eq:H-Y_depolarization} 
	\end{align}
	
	There are several SPH depolarization channels that can possibly be present in this reaction. Collisions between the $^2S_{1/2}$ SPH atoms and $^2P_{3/2}$ Y atoms are predominantly inelastic, and governed by the spin-orbit interaction \cite{Happer_1972,Bouchiat_1969}. These inelastic collisions result in the randomization of the SPH electron polarization in two possible ways: either via the formation of a short-lived molecular state, where the SPH electron spin couples to the orbital angular momentum of the resulting molecule, and is therefore lost when the molecule is dissolved (e.g. by another collision): 
	
	\begin{align}
		& SPH + Y \rightarrow H\mspace{-8.0mu}-\mspace{-7.0mu}Y \rightarrow H + Y\label{eq:sticking} 
	\end{align}
	or by "sudden" collisions, which occur do not form a molecular state, but have a finite probability to destroy the SPH electron polarization upon impact:
	
	\begin{align}
		& SPH + Y \rightarrow H + Y\label{eq:sudden} 
	\end{align}
	
	The former type of collisions, known as "sticking" collisions, play a more important role at lower pressures in the mbar range \cite{Bouchiat_1969}. We can therefore expect that SPH depolarization is governed by "sudden" H-Y collisions. The depolarization cross-section in that case is given by $\sigma_{H-Y} = \alpha \sigma_{coll}$, where $\sigma_{coll}$ is the hard-sphere collision cross section for hydrogen-halogen collisions, and $\alpha$ is a parameter giving the probability of disorientation of the SPH electron polarization upon impact, with a value that is typically much less than 1. The parameter $\alpha$ depends on the interaction strength and duration ($\approx 10^{-12}$ for the spin-orbit interaction). As $\sigma_{coll}$ is $\sim1.5\times10^{-15}$ \si{\centi\meter\squared}, we expect $\sigma_{H-Y}$ to be of order $10^{-16}-10^{-17}$ \si{\centi\meter\squared}. To our knowledge, an exact value of the depolarization cross-section of SPH by halogen radicals is not cited in the literature. 
	
	Before proceeding to the determination of the depolarization cross-section, let us note that there are several other possible SPH reaction channels in the configuration discussed in this study, which, however, are ignored, as they are considered negligible compared to the sudden, binary SPH-Y collisions. A spin-exchange interaction-induced relaxation channel is in principle possible, but has never been observed in collisions between S and P atoms, and has to be considered minor compared to the main relaxation mechanism, which is governed by the spin-orbit interaction. Another possible relaxation channel could be the reaction of SPH with a halogen diatomic molecule, which could be formed via three-body recombination of two halogen atoms:
	
	\begin{align}
		& \label{3Y_body} 2Y + M \rightarrow Y_{2}+M\\
		& \label{eq:HY2ch2_body} SPH + Y_{2} \rightarrow HY + Y
	\end{align}	
	however, this reaction can be ignored without the presence of a high density of a third body M that can efficiently catalyze the halogen recombination reaction, e.g. an inert gas(see appendix).
	
	\section{Determination of the depolarization rate}\label{sec:experiment}
	
	
	Sofikitis \textit{et al.} \cite{Sofikitis_HighDensity} used a simple method to estimate $\sigma_{H-Y}$. They focused the beam with two lenses, one with f=25 mm and one with f=50 mm, placed at a distance $l\sim25 mm$ from the center of the coil.
	The weaker lens focused the beam outside the coil, and produced an SPH density of $\sim10^{16}$\si{\per\cubic\centi\meter} inside the coil, whereas the stronger lens focused the beam at the center of the coil, creating an effective SPH density of $\sim10^{19}$ \si{\per\cubic\centi\meter} near the focus. This way, they expected to differentiate between the "low"- and "high"-density depolarization rates. However, they observed lifetimes of $\sim10-20$ ns under both density regimes, yielding no evidence of SPD depolarization from $I({}^{2}P_{3/2})$ in the "high"-density regime.
	The deuterium iodide (DI) density could not be increased above $\sim10^{19}$ \si{\per\cubic\centi\meter}, to increase the depolarization rate, because the absorption cross section of DI at 266 nm, $2\times10^{-19}$ \si{\square\centi\meter}, is large enough to prevent sufficient photolysis laser light reaching the laser focus, for the geometry of the experiments.
	The absorption cross section of HCl at 213 nm is 2 orders of magnitude lower, at $2\times 10^{-21}$\si{\square\centi\meter}. Therefore, the HCl density can be increased to $10^{20}$ \si{\per\cubic\centi\meter}, and the photolysis laser can reach the laser focus in the cell, without too much loss from absorption, as the optical depth is $\sim$5 cm. Therefore, we are able to study the SPH depolarization dynamics at significantly higher densities, needed to determine the SPH-Cl depolarization cross section.
	
	\subsection{Experiment}\label{sec:ch2Experiment}
	
	\begin{figure*}[t!]
		\includegraphics[width=0.9\textwidth]{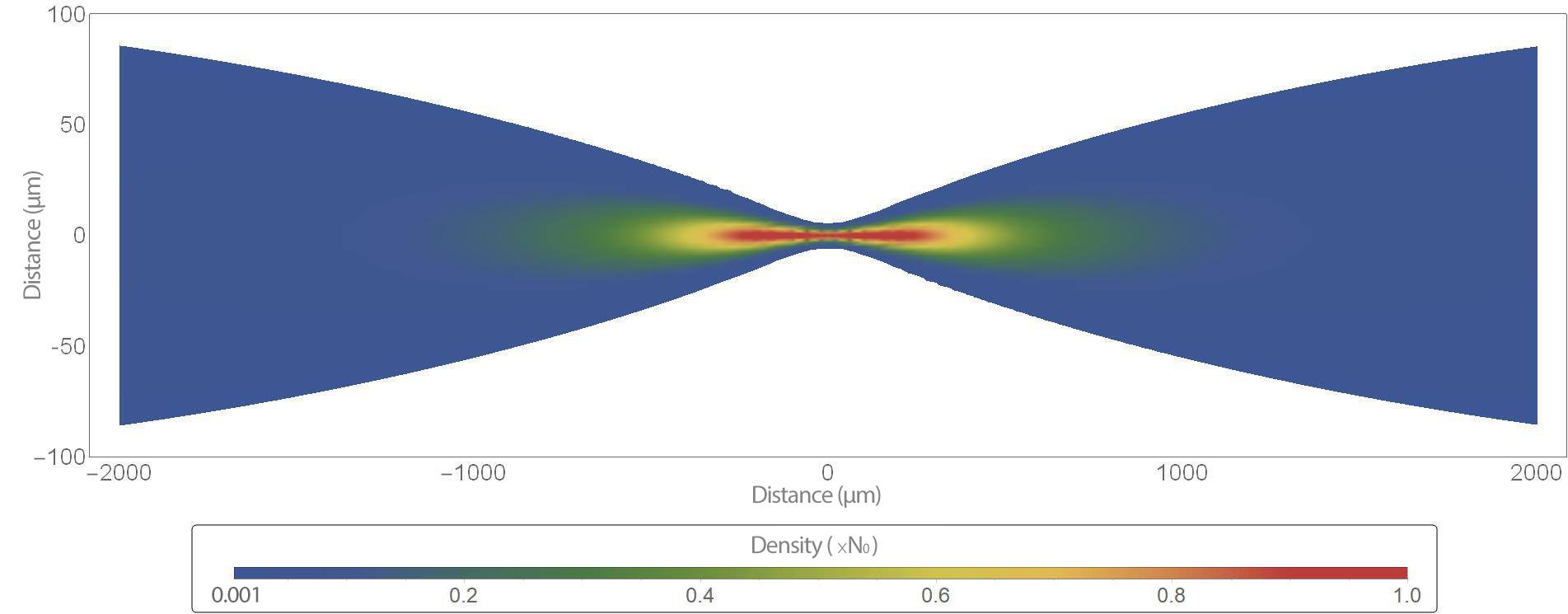}
		\caption{\label{fig:densityplot} Plot of the SPH density inside the coil, after the photodissociation of 2 bar HCl with number density $N_0=5.4\times10^{-19}$\si{\per\centi\meter\cubed} by a 3 mJ, 213 nm pulse, focused by a lens with f=5 \si{\centi\meter}.  Note that the lower limit of the plot densities is 0.1\% (blue), and it increases to 100\% (red) at the laser focus.}
	\end{figure*}
	
	A simple experiment was designed to estimate $k_{H-Y}$, by measuring
	the SPH polarization lifetime for various SPH densities. The experimental setup is the same as the one used in \cite{Sofikitis_HighDensity}, with hydrogen chloride (HCl) as the SPH source gas.
	Briefly, 1-5 bar HCl are introduced into the gas cell, and a circularly polarized, 213 nm, 150 ps laser pulse is used to photodissociate the HCl. The 4 mm diameter beam is focused by a 5 cm lens, placed on a one-dimensional translation stage, by which we control the distance of the lens from the cell, and therefore the position of the focus inside the cell. A pickup coil with a length of 4 mm, 4.5 turns, and a 2 mm diameter, is used to detect the time evolution of the SPH produced by the photodissociation of HCl. The time-dependent signal that the coil picks up is of the form:
	
	\begin{equation}\label{envelope}
		I(t) = e^{-t/\tau}\cos{2 \pi f t}
	\end{equation}
	
	Where t is the time, $\tau$ the polarization lifetime, and f is the hyperfine beating frequency, which is equal to 1.420415 GHz for the unperturbed hyperfine hydrogen Hamiltonian 
	\begin{equation}\label{eq:Hamiltonianch2}
		H = A_{HF} \textbf{I} \cdot \textbf{J}
	\end{equation}
	with $A_{HF}$ the hyperfine constant, and \textbf{I}, \textbf{J} the quantum operators for the nuclear and electron spin, respectively.
	With this setup, we can scan the position of the focus through the coil length, effectively varying the density of the SPH produced inside the coil.
	
	\subsection{Calculation of the Expected SPH Density}\label{sec:SPH_calc}
	
	\subsubsection{Rate equations}
	From the discussion in section \ref{sec:mechanism}, we can take the rate equations of the relaxation of SPH from collisions with Y radicals: 
	
	\begin{subequations}\label{SPH_rateeq_no_recombination}
		\begin{eqnarray}
			&\frac{d[SPH]_t}{dt} = - k_{H-Y} [SPH]_t [Y]_t -
			k_{H-HY} [SPH]_t [HY]_t\\
			&\frac{d[Y]_t}{dt} = 0\\
			&\frac{d[HY]_t}{dt} = 0
		\end{eqnarray}
	\end{subequations}
	where we make the approximation that no long-lived molecular states are formed, as our experiments were conducted at high gas densities. It follows that, to calculate $k_{H-Y}$, the SPH (and Y) density immediately after the photodissociation must be known. If the photodissociation is 100\% efficient, $HY(0)=0$, and the second term of \ref{SPH_rateeq_no_recombination}a vanishes. Otherwise, the second term is non-negligible, and can be calculated using the value of $k_{H-HY}$ from \cite{Boulogiannis_2019} ($7.5\times10^{-19}$\si{\centi\meter\squared}). 
	
	Note that, as mentioned in \cite{Sofikitis_HighDensity}, and was confirmed in this study, no evidence of spin-polarized halogens (i.e., a Fourier peak at 150 MHz, the hyperfine frequency of Cl ${ }^2P_{3/2}$ \cite{Chlorine_HF}) is found in the data. The absence of such evidence means that halogen atoms are quickly ($\ll$1 ns) depolarized. This supports the conclusion that the main depolarizer of SPH is the unpolarized halogen atom; if the halogen atom retained its polarization for a longer time, the H-Y relaxation rate would be much lower.
	
	\subsubsection{Spatial distribution of SPH density}
	
	The dimension of the beam was confirmed to follow the distribution of a Gaussian beam using a razor-edge scheme. The waist at focus was estimated
	to be 6$\pm1 \mu$m, close to the diffraction limit of d = 5 $\mu$m. A theoretical model based on the Beer-Lambert law, taking saturation into account, is used to simulate the densities created by a laser beam of known dimensions and energy. To account for saturation, i.e. for when the photon density is high enough that virtually all gas atoms are photodissociated, we use the`following considerations.

	The density of molecules available for dissociation in each position inside the volume occupied by the laser beam depends on the local laser intensity, i.e. the laser pulse energy and cross sectional area. In our one dimensional model, both of these quantities depend on z. The variation of the laser waist w(z) as a function of distance z is given by\cite{verdeyen}: 
	\begin{equation}\label{eq:w_z}
		w(z)= \sqrt{\frac{\lambda z_{01}}{\pi} \big[(1-z/f)^2+(z/z_{01} )^2 \big]}
	\end{equation}
with $z_{01}=\frac{\pi w_{01}^2}{\lambda}$, and where $w_{01}$ is the laser beam waist at the surface of the focusing lens. The pulse energy is reduced as the pulse propagates through the sample due to absorption.

The incremental change of the density of molecules available for dissociation, $dN$, over the incremental change in the number of dissociation photons, $dn$, is proportional to the density of molecules, $N$, the dissociation cross section, $\sigma_{HCl}$ and inversely proportional to the pulse area $\alpha$, as  $\frac{dN(z)}{dn}=-\sigma_{HCl} \frac{N (z)}{a(z)}$ 
resulting in a relation for the number of molecules available for dissociation as a function of the distance z inside the coil, $N(z)= N_0 e^{-\sigma_{HCl}\frac{n}{a(z)}}$.
If we include this dependence of N(z) in the usual Beer-Lambert law, we end up with the equation:
	\begin{equation}\label{eq:BL_sat}
		\frac{dE(z)}{dz} = -\sigma_{HCl} N_0 E(z)\times exp\Big[-\sigma_{HCl}  \frac{E(z)}{a(z) E_{ph}}\Big]
	\end{equation}
	where $E(z)$ is the pulse energy at a distance z from the laser source, $\sigma_{HCl}$ the absorption cross section of HCl, $N_0$ the initial HCl density, $E_{ph}$ the single photon energy, and \textit{a} the laser beam area cross-section. This equation simplifies to the Beer-Lambert law when the photon density is much lower than the absorption cross-section, i.e. there is no optical saturation, and gives a plateau in absorption when optical saturation occurs.
	%
	%
	
	The produced SPH density through the length pickup coil length, is shown in figure \ref{fig:densityplot}, for an HCl pressure P=2 bar, corresponding to an initial number density of $5.4\times10^{19} cm^{-3}$. The model shows that the SPH density within the Rayleigh range of the beam ($z_o\sim0.5 mm$) is close to the initial HCl density for pressures of up to 2 bar.
	
	However, due to the large beam divergence and the low absorption cross-section of hydrogen chloride, the density outside of the Rayleigh range is much lower, 
	of order $10^{17-18} cm^{-3}$. If SPH depolarization at a high density environment is fast, this variation in density within a few mm would result in polarization lifetime gradients in space. Since the pickup coil is longer than this length, the different SPH polarization lifetimes would be imprinted in the pickup coil signal (\ref{envelope}) with a position-dependent exponential:
	\begin{equation}\label{envelope_sumexp}
		I(t) = 
		\iint_V \rho_{SPH}(r,\theta) e^{-t/\tau(r,\theta)}\cos{(2 \pi f t)}\,dr\,d\theta
	\end{equation}
	where $\rho_{SPH}$ and $\tau$ depend on r and $\theta$.
	
	A shorter coil would, of course, offer better spatial resolution, but at the detriment of the signal-to-noise ratio(SNR). For this reason, in this experiment, we chose to work with a 4.5 turns, 4 mm long pickup coil. Further investigation on the optimization of the coil is needed, to achieve shorter length while retaining a high SNR.
	
	The laser used in this experiment emits a $\lambda$ = 213 nm, $\tau_p$ = 150 ps, E = 3 mJ, D = 4 mm Gaussian pulse. The pick-up coil’s diameter and length are d = 2 mm and l = 4.5 mm respectively, and is placed 3 cm away from the window of the cell. The coil is relatively short in order to facilitate the acquisition of rapidly changing signal. The photodissociation cross section for HCl at 213 nm (room temperature) is $\sigma = 1.7 \times 10^{-21}$ \si{\squared\centi\metre} \cite{HCl_Cross_Section}.
	
	From figure \ref{fig:densityplot}, obtained via eq. \ref{eq:BL_sat} for the aforementioned conditions, we can observe the variation in SPH density inside the coil. Inside the Rayleigh range of the beam, HCl is almost or entirely depleted (70-100\%), and there exist only hot spin-polarized hydrogens and chlorines immediately after the photodissociation. As discussed earlier, Cl is almost entirely depolarized within less than 1 ns after the photodissociation. The fast depolarization would then be triggered by collisions between SPH and unpolarized Cl. At positions outside the Rayleigh range, SPH and Cl account for less than 1\% of the total number density, and are surrounded by HCl molecules. In that case, the main depolarization mechanism is via the HCl-SPH complex formation (\ref{complex_formation}), and the depolarization rate is more than an order of magnitude lower. 
	Using this density calculation for H, Cl, and HCl at every position of the beam, we can extract a value for the H-Cl depolarization cross section from eqs. \ref{SPH_rateeq_no_recombination}. To that end, a Finite Element Analysis (FEA) is employed, where the volume of the beam is separated in sections of \{x,y,z\}=1 \si{\micro\meter} $\times$ 1\si{\micro\meter} $\times$ 20\si{\micro\meter}, and the time evolution of the SPH in each section is calculated, by solving eqs. \ref{SPH_rateeq_no_recombination} in each volume. It follows that, in volumes with "low" SPH density, where $[Y]\ll[HY]$, the second term of \ref{SPH_rateeq_no_recombination}a is dominant, while in volumes with "high" density,where $[Y]\gg[HY]$, the depolarization rate is dominated by the first term of \ref{SPH_rateeq_no_recombination}a. The latter condition applies, as we can see from figure \ref{fig:densityplot} to the Rayleigh range of the beam, whereas immediately outside the Rayleigh range, the densities of SPH, Y and HCl are comparable, until the HCl density dominates farther away from the Rayleigh range of the beam.
	
	\subsection{Distribution of Velocities}
	
	To extract $\sigma_{H-Cl}$, the distribution of velocities should also be calculated. The velocity of SPH after HY photodissociation is $\tilde{v}\approx$20 km/s \cite{Sofikitis_HClHBr_SPH}. However, thermal equilibrium between H, Cl (and HCl, where less than 100\% of HCl molecules have been photodissociated) should be reached quickly after the photodissociation, as the hard-sphere collision rate is 40 \si{\per\nano\second}. 
	
	Using the Equipartition theorem, 
	\begin{equation}\label{eq:Equip_theorem}
		E_{EKE} = x E_H + x E_Cl + (1-x) E_{HCl}
	\end{equation}
	
	where $E_{EKE}=E_{photon}-E_{bond}=1.4 eV$ is the excess kinetic energy of hydrogen after the photodissociation, $E_H, E_{Cl}$ and $E_{HCl}$ the energies of hydrogen atoms, chlorine atoms, and hydrogen chloride molecules at equilibrium, respectively, and x = n/$N_0$ the fraction of photodissociated molecules, with n the SPH density produced by the photodissociation, and $N_0$ the number density of HCl prior to the photodissociation. Here, we assumed the kinetic energy of chlorine atoms and hydrogen chloride molecules immediately after the photodissociation to be small compared to that of hydrogen. Solving eq. \ref{eq:Equip_theorem}, the total kinetic energy of the H atoms at thermal equilibrium, $E_{H_{tot}}$, as a function of the dissociated fraction x of the HCl molecules, is given by:
	
	\begin{equation}\label{eq:Energy after}
		E_{H_{tot}} = E_{H_{300K}}+ E_{EKE} \times \big[\frac{3x/2}{3 x + \frac{C_{p_{HCl}}}{R} (1-x)} \big]\\
	\end{equation}
	where  $C_{p_{HCl}}$ is the heat capacity of HCl, R is the ideal gas constant, and $E_{H_{300K}}$= 0.038 \si{\electronvolt} is the thermal energy of hydrogen at room temperature, which is added to the excess kinetic energy, to obtain the total hydrogen energy at equilibrium.
	The black line in figure \ref{fig:eq_velocity} shows the equilibrium SPH velocity as a function of the SPH density. We see that the thermal equilibrium velocity of hydrogen when HCl is completely photodissociated is reduced to $\sim$12 km/s, as a result of the frequent collisions with the initially much colder($\sim$300 K) Cl atoms, whereas at low SPH densities, SPH velocity is the room temperature velocity, $\sim$2.5 km/s, determined by the room temperature HCl. 
	
	\section{Results and Discussion}\label{sec:results}

	\subsection{Results}
\begin{figure*}[h!]
	\includegraphics[width=0.9\textwidth]{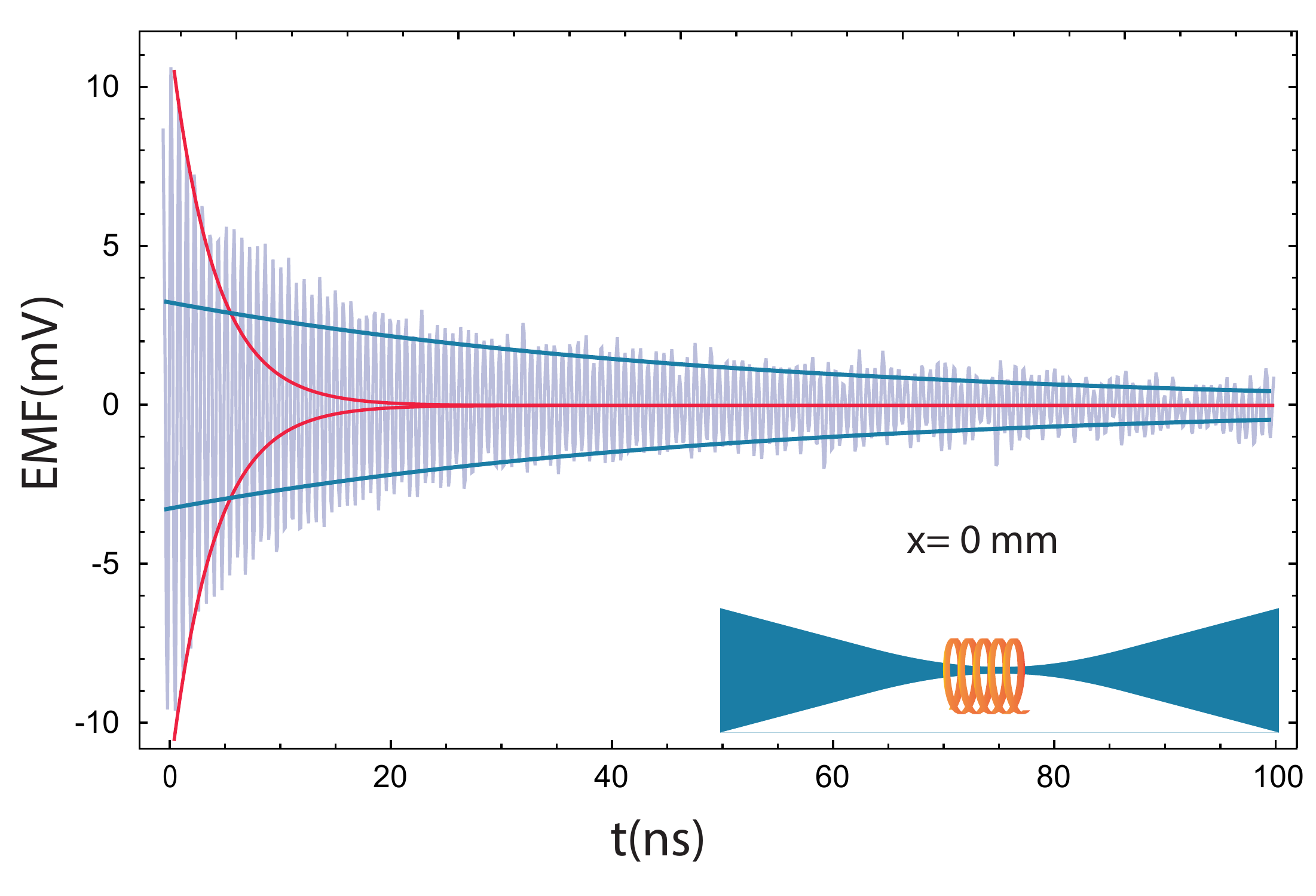}
	\caption{\label{fig:single_Trace} Experimental Data for $P_{HCl}$=2 bar, with the laser beam focused at the center of the coil (Light Blue). The data cannot be fit with a single exponential. We show approximate exponential fits for the early part of the trace with $\tau$=5 \si{\nano\second} (Red), and for the late part of the trace with $\tau$=50 \si{\nano\second} (Petrol Blue).}
\end{figure*}
	We can now use the H, Cl, and HCl density distribution, the SPH-Cl and SPH-HCl relative velocities distribution in eq. \ref{SPH_rateeq_no_recombination} to determine $\sigma_{H-Cl}$. Figure \ref{fig:single_Trace} shows a typical dataset at an HCl pressure P=2 bar, when the pulse is focused by a 5 cm lens at the center of the coil. The experimental traces demonstrates two distinct segments. In the first $\sim$ 5 ns, there is a steep drop observed in the early part of the signal, followed by a lower slope at later times. We attribute the steep drop to the high density inside the Rayleigh range, where the main depolarization mechanism is via electron randomization by collisions between SPH and Cl. At later times, the slope becomes gradually lower, approaching the low-density depolarization rate, which was measured by Boulogiannis et al.\cite{Boulogiannis_2019}. As shown in fig.  \ref{fig:single_Trace}, this signal shape cannot be fitted by a single exponential, because the depolarization rate depends on the density of the produced SPH at a specific position, and therefore has a spatial dependence. Thus a finite element analysis is a more suitable  method for the interpretation of the data, and the extraction of a depolarization cross section for SPH-Cl collisions. 
	
\begin{figure*}[h!]
		\includegraphics[width=0.9\textwidth]{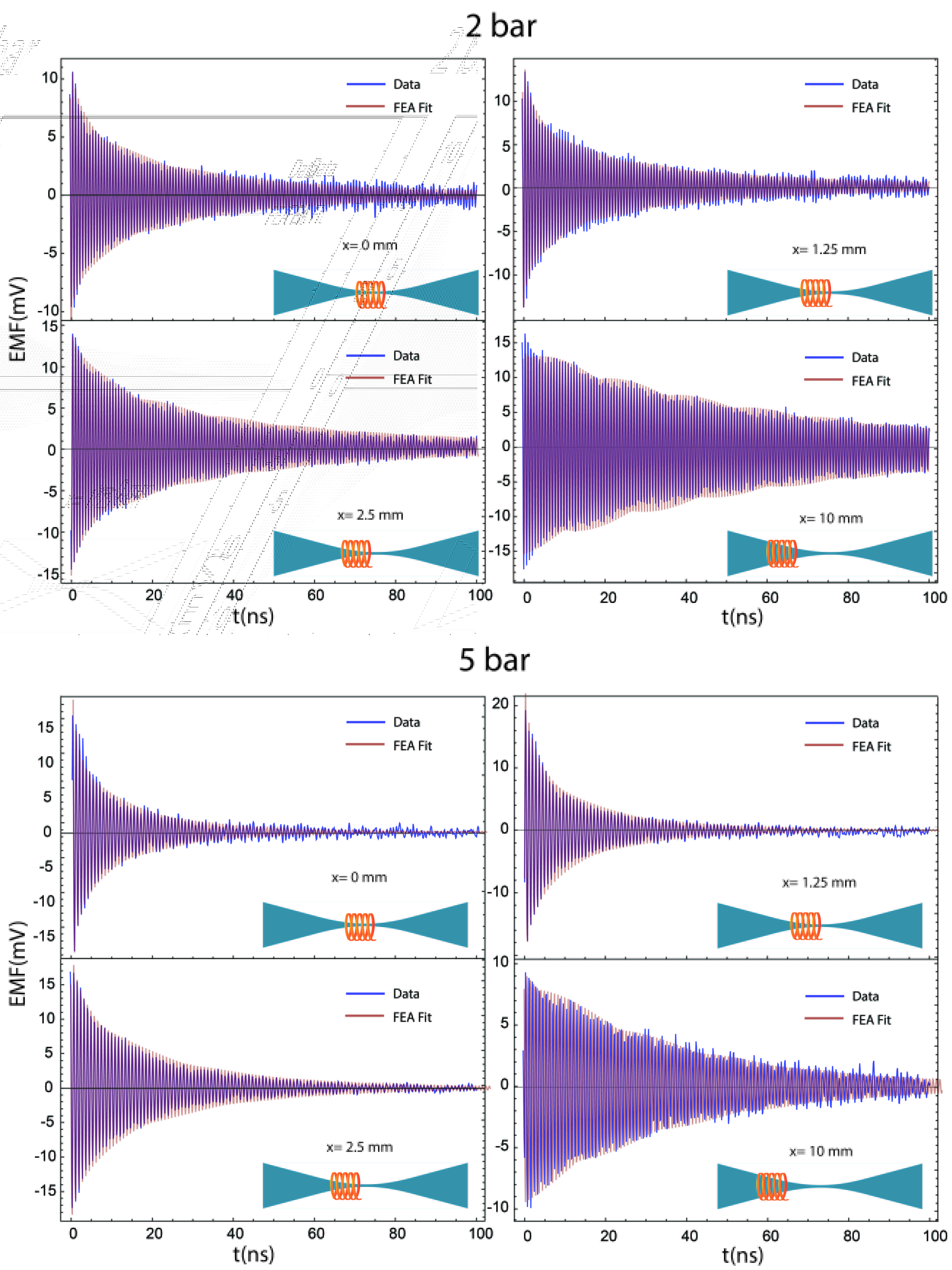}
		\caption{\label{fig:correct_sigmas} Experimental Data (Blue) and fits generated by finite element analysis of the produced SPH inside the pickup coil (Light Red) for different positions of the beam focus related to the coil center, at an initial HCl pressure of 2 bar (upper panel) and 5 bar (lower panel). Fits are shown for $\sigma_{H-Cl}=7\times10^{-17}$\si{\centi\meter\squared}, and x is the distance of the lens relative to the position where the beam is focused at the center of the coil.}
	\end{figure*}

	Figure \ref{fig:correct_sigmas} shows the SPH signal after photodissociation by the laser pulse, and a fit with $\sigma_{H-Cl}=7\times10^{-17}$\si{\square\centi\meter}, for various focusing geometries, and at two initial HCl pressures, 2 and 5 bar. The value of $\sigma_{H-Cl}=7\times10^{-17}$ \si{\square\centi\meter} fits the data well at all focusing geometries and at both 2 bar and 5 bar, lending confidence to the determination of the SPH relaxation cross section. We gain confidence in our model, based on Eqs. \ref{eq:BL_sat} and \ref{eq:Energy after}, because the experimental data (and their deviations from exponential decays) can be fit very well with a single cross section (i.e. the model seems to correctly calculate the SPH, Cl, and HCl density dependence on position).
	Taking the noise of the experimental trace into account, we can determine that the value of the depolarization rate of SPH produced via photodissociation of high densities of HCl is $\sigma_{H-Cl}=7(2)\times10^{-17}$ \si{\centi\meter\squared}.
	
\begin{figure*}[h!]
		\includegraphics[width=0.9\textwidth]{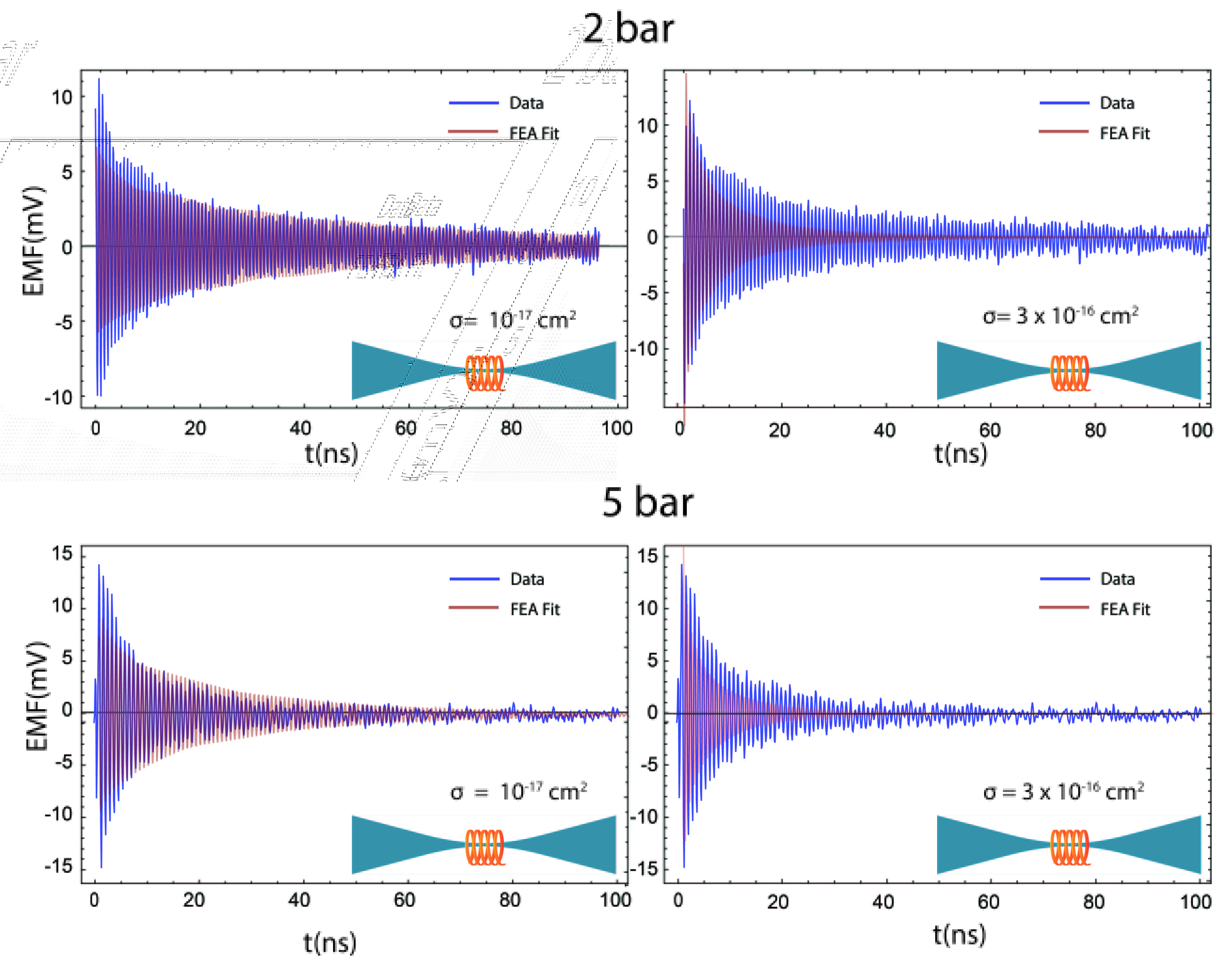}
		\caption{\label{fig:wrong_sigmas}Same as fig. \ref{fig:correct_sigmas}, with the beam focused inside the coil, and two different values for the H-Cl cross section, $\sigma_{H-Cl}=10^{-17}$ \si{\centi\meter\squared} (left) and $\sigma_{H-Cl}=3 \times 10^{-16}$\si{\centi\meter\squared} (right), showing clearly that these two cross sections are too small and too large, respectively.}
	\end{figure*}

	It is instructive to demonstrate the simulations produced by our model, when the value of the cross section is fixed to be much higher, or much lower than the one that fits the experimental data. Simulations with different values of the depolarization cross-section are shown in figure \ref{fig:wrong_sigmas}, where we can see that values of $10^{-17}$ and $3\times10^{-16}$ provide poor fits to the data. Specifically, a low $\sigma_{H-Cl}$ does not fit the data well at short times, and resembles a single-exponential fit that only fits the low-density regime well; whereas, a much larger value of $\sigma_{H-Cl}=3\times10^{-16}$ \si{\centi\meter\squared} emulates a fast single exponential that only approaches the early part of the data. 
	
	\subsection{Use of an inert gas for the suppression of the depolarization}
	
	In this section, we discuss the potential effect of a buffer gas in the depolarization rate of SPH and SPD. An inert gas with a high heat capacity can be used to cool the SPH down to a lower equilibrium temperature, and as such lower the collision rate, and thus lower the depolarization rate (assuming that the depolarization cross section does not increase too much at low energies). 
	
	A suitable candidate may be hexafluorethane ($C_2F_6$), which has a heat capacity of 105 \si{\joule\per\mole\per\kelvin} at room temperature, which increases to over 170 \si{\joule\per\mole\per\kelvin} at high temperatures. Additionally, $C_2F_6$ is transparent at middle and far ultraviolet, reducing the possibility of reacting fragments emerging from dissociation or ionization. 
	
	To account for the effect of $C_2F_6$, we add an additional term k$E_{C_2F_6}$ in eq. \ref{eq:Equip_theorem}, where $k=N_{0_{C_2F_6}}/N_{0HCl}$ the number density ratio of $C_2F_6:HCl$, and $E_{C_2F_6}$ the equilibrium thermal energy of SPH. Then, eq. \ref{eq:Energy after} becomes:
	
	\begin{equation}\label{eq:Energy after C2F6}
		E_{H_a} = E_{H_{300K}}+ E_{EKE} \times \frac{3x/2}{3 x + \frac{C_{p_{HCl}}}{C_{p_{H}}} (1-x) + k \frac{C_{p_{C_2F_6}}}{C_{p_{H}}}} 
	\end{equation}
	
\begin{figure*}
		\includegraphics[width=\textwidth]{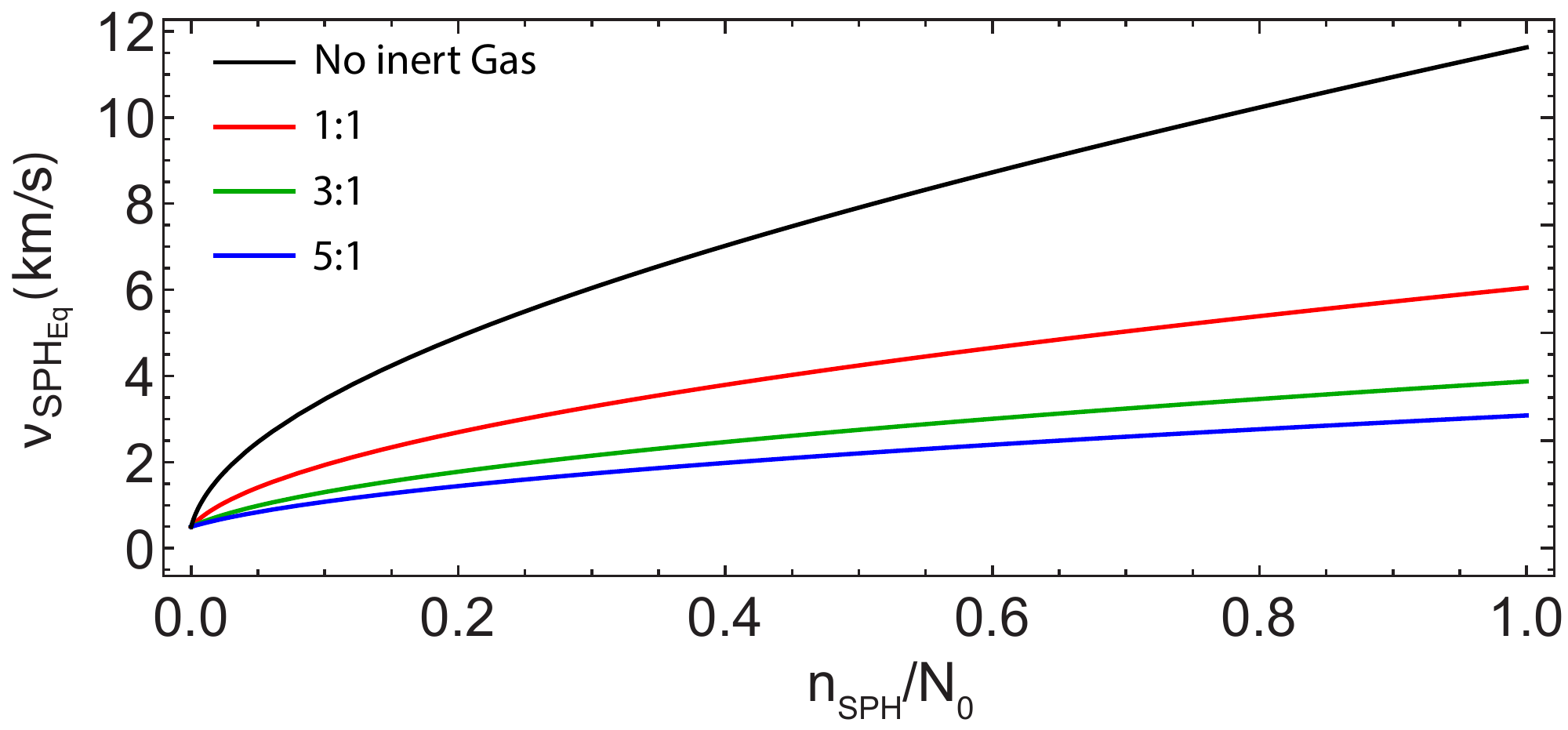}
		\caption{\label{fig:eq_velocity}SPH thermal velocity distribution vs HCl photodissociation fraction, for various ratios of $C_2F_6$ partial pressure to HCl partial pressure.}
	\end{figure*}

	The equilibrium velocity of SPH after HCl photodissociation in the presence of  various partial pressures of $C_2F_6$ is shown in figure \ref{fig:eq_velocity}. The reduction of the equilibrium velocity by a factor of 3 observed with a 3:1 $C_2F_6$/HCl ratio is expected to prolong the polarization lifetime by an equal factor at the high-density regime.
	
	\subsection{Polarization of electrons, protons and deuterons}
	
	While the degree of electronic polarization of SPH can be practically 100\%, as electronic polarization production is nearly instantaneous, the polarization of the nucleus occurs after a hyperfine half-period, which is $\sim$350 ps for SPH and $\sim$1.52 ns for SPD. Therefore, the degree of polarization of the SPH nucleus depends on the polarization lifetime.
	
\begin{figure*}
		\includegraphics[width=0.9\textwidth]{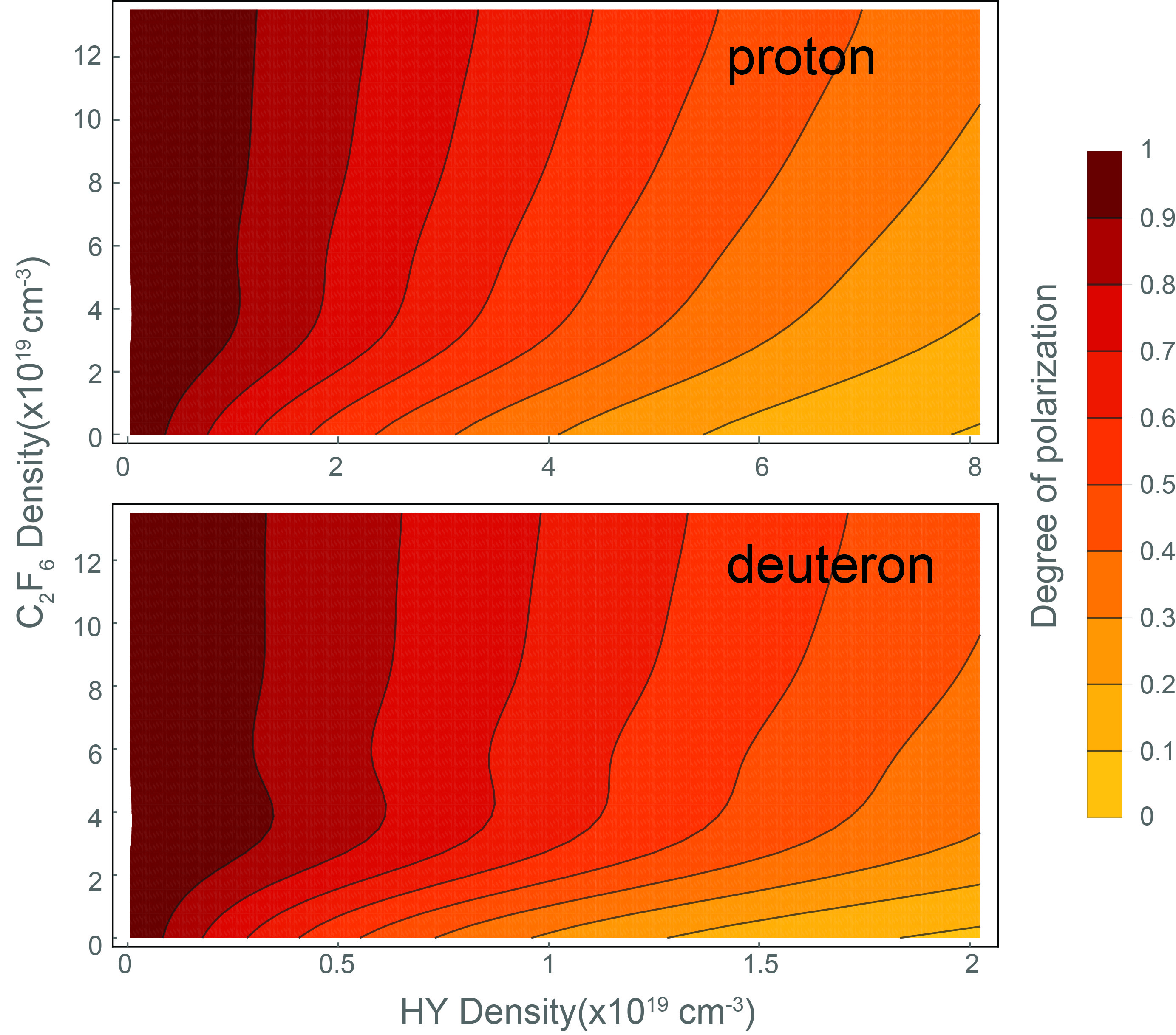}
		\caption{\label{fig:density_limit} Degree of polarization of protons, 350 ps after complete photodissociation of HCl(upper figure), and deuterons, 1.5 ns after complete photodissociation of DI(lower figure), at various pressure conditions, and at an initial temperature of T=300 K.}
	\end{figure*}

	In fig. \ref{fig:density_limit}, we see a simulation of the expected degree of polarization of spin-polarized protons, after 100\% photodissociation of HCl with a 213 nm pulse, and after a time delay of 350 ps (for the polarization transfer from the electron to the proton) (\ref{fig:density_limit}a), and of spin-polarized deuterons, after 100\% photodissociation of DI, and after a time delay of 1.5 ns (for the polarization transfer from the electron to the deuteron) (\ref{fig:density_limit}b), in the presence of  $C_2F_6$. For these simulations, we assumed $\sigma_{H-Cl}=\sigma_{D-I}=7\times10^{-17}$, the experimentally determined value for the SPH cross section at the "high" density regime. A notable result is that $\ge$ 90\% polarized protons can be produced at a density of about $10^{19}$\si{\per\centi\meter\cubed}, and the same degree of polarization for deuterons is achievable up to $\sim3\times10^{18}$ \si{\per\centi\meter\cubed}. A threefold increase in proton densities would yield a polarization of 70\%, equal to that for deuteron densities of $10^{19}$ \si{\per\centi\meter\cubed}. 
	
	We can then deduce expected values for spin polarized electron, proton and deuteron densities produced via the photodissociation method. About $2.7\times10^{20}$ \si{\per\centi\meter\cubed} 100\% spin-polarized electrons can be produced via complete photodissociation of 5 bar HCl, followed by ionization of the SPH with 243 nm light. Similarly, $>$70\% spin-polarized protons and deuterons can be produced in the same way, only with ionization occuring after a hyperfine beating half-period (350 ps for SPH, and 1.5 ns for SPD), at densities of $\sim3\times10^{19}$ \si{\per\centi\meter\cubed} for protons and  $\sim10^{19}$ \si{\per\centi\meter\cubed} for deuterons. 
	
	If the depolarization cross section at low temperatures is determined to
	be sufficiently low, it will then be worth cooling the gas in a supersonic
	expansion to a few K, so that the SPH or SPD can thus be translationally
	cooled to a few K, which may lower the SPH depolarization rate even more,
	and allow even higher densities. Further studies into the collision-energy
	dependence of the depolarization cross section are needed, to determine
	the optimal gas temperature of maximum SPH/SPD densities.

	\section{Conclusions}
	To our knowledge, the value of $\sigma_{H-Cl}$ presented here is the first experimentally measured value of the SPH relaxation rate with Cl atoms, and is consistent  with the upper limit of $\sigma_{D-I}$ that Sofikitis \textit{et al.} \cite{Sofikitis_HighDensity} suggested in an experiment conducted at one order of magnitude lower densities for DI photodissociation. The measured value shows that lifetimes of order of a few ns are possible, for SPH at densities of about $10^{19}$ \si{\per\centi\meter\cubed}. Calculation of $\sigma_{H-Cl}$, particularly as a function of collision energy, will be helpful to corroborate the results and conclusions of this work, or to help point out issues that still need to be elucidated.

	The experimental method presented here could be improved to test higher SPH densities with a better spatial resolution, by using a shorter coil. This, however, would reduce the inductance and quality factor, and consequently the signal-to-noise ratio. A potential solution for this would be the use of a microstrip coil, which offers a high quality factor at GHz frequencies at a very compact size. Furthermore, higher energy lasers would produce higher SPH densities, possibly without the need for tight focusing, thus eliminating the SPH distribution gradients that the 3 mJ laser used in this study creates.
	Ultrahigh density SPH has several novel applications, including ns-resolved magnetometry, laser-ion acceleration, and tests of polarized nuclear fusion \cite{Spiliotis_LSA}.
	
The experimental work was conducted at the Ultraviolet Laser Facility at FORTH-IESL, supported in part by the Hellenic Foundation for Research and Innovation (HFRI) and the General Secretariat for Research and Technology (GSRT), through the grant agreement No. HFRI-FM17-3709 (project NUPOL). We thank S. Pissadakis for access to the EKSPLA SL312M laser. 
	 
	\clearpage
	\bibliography{SPH_Cl_depolarization}

\begin{thebibliography}{10}

\bibitem{Szczerba_2000}
D~Szczerba, L.D {van Buuren}, J.F.J {van den Brand}, H.J Bulten, M~Ferro-Luzzi,
  S~Klous, H~Kolster, J~Lang, F~Mul, H.R Poolman, and M.C Simani.
\newblock A polarized hydrogen/deuterium atomic beam source for internal target
  experiments.
\newblock {\em Nuclear Instruments and Methods in Physics Research Section A:
  Accelerators, Spectrometers, Detectors and Associated Equipment}, 455(3):769
  -- 781, 2000.

\bibitem{Clasie_2006}
B.~Clasie, C.~Crawford, J.~Seely, W.~Xu, D.~Dutta, and H.~Gao.
\newblock Laser-driven target of high-density nuclear-polarized hydrogen gas.
\newblock {\em Phys. Rev. A}, 73:020703, Feb 2006.

\bibitem{SEOP_SPH}
S.~G. Redsun, R.~J. Knize, G.~D. Cates, and W.~Happer.
\newblock Production of highly spin-polarized atomic hydrogen and deuterium by
  spin-exchange optical pumping.
\newblock {\em Phys. Rev. A}, 42:1293--1301, Aug 1990.

\bibitem{Walker_1997}
Thad~G. Walker and William Happer.
\newblock Spin-exchange optical pumping of noble-gas nuclei.
\newblock {\em Rev. Mod. Phys.}, 69:629--642, Apr 1997.

\bibitem{Sofikitis_HighDensity}
Dimitris Sofikitis, Chrysovalantis~S. Kannis, Gregoris~K. Boulogiannis, and
  T.~Peter Rakitzis.
\newblock Ultrahigh-density spin-polarized {H} and {D} observed via
  magnetization quantum beats.
\newblock {\em Phys. Rev. Lett.}, 121:083001, Aug 2018.

\bibitem{Boulogiannis_2019}
Gregoris Boulogiannis, Chrysovalantis Kannis, Georgios Katsoprinakis, Dimitrios
  Sofikitis, and T.~Rakitzis.
\newblock Spin polarized hydrogen depolarization rates at high hydrogen-halide
  pressures: Hyperfine depolarization via the hy-h complex.
\newblock {\em The Journal of Physical Chemistry A}, 123, 09 2019.

\bibitem{Spiliotis_Magnetometry}
Alexandros~K. Spiliotis, Michael Xygkis, Konstantinos Tazes, Michael Koutrakis,
  George.~E. Katsoprinakis, Georgios Vasilakis, and T.~Peter Rakitzis.
\newblock A nanosecond-resolved atomic hydrogen magnetometer.
\newblock {\em arXiv:2010.14579}, 2020.

\bibitem{Wu_2019}
Yitong Wu, Liangliang Ji, Xuesong Geng, Qin Yu, Nengwen Wang, Bo~Feng, Zhao
  Guo, Weiqing Wang, Chengyu Qin, Xue Yan, Lingang Zhang, Johannes Thomas, Anna
  Hützen, Markus Büscher, T~Peter Rakitzis, Alexander Pukhov, Baifei Shen,
  and Ruxin Li.
\newblock Polarized electron-beam acceleration driven by vortex laser pulses.
\newblock {\em New Journal of Physics}, 21(7):073052, jul 2019.

\bibitem{Wen_2019}
Meng Wen, Matteo Tamburini, and Christoph~H. Keitel.
\newblock Polarized laser-wakefield-accelerated kiloampere electron beams.
\newblock {\em Phys. Rev. Lett.}, 122:214801, May 2019.

\bibitem{Jin_2020}
Luling Jin, Meng Wen, Xiaomei Zhang, Anna Hützen, Johannes Thomas,
  M.~Büscher, and Baifei Shen.
\newblock Spin polarized proton beam generation from gas-jet targets by intense
  laser pulses.
\newblock {\em Physical Review E}, 102:011201, 01 2020.

\bibitem{Buscher_2020}
M.~et~al Büscher.
\newblock Simulation of polarized beams from laser-plasma accelerators.
\newblock {\em Journal of Physics Conference Series}, 1596:012013, 09 2020.

\bibitem{Sofikitis_SPD_PRL}
Dimitrios Sofikitis, Pavle Glodic, Greta Koumarianou, Hongyan Jiang, Lykourgos
  Bougas, Petros Samartzis, Alexander Andreev, and T.~Rakitzis.
\newblock Highly spin-polarized deuterium atoms from the {UV} dissociation of
  deuterium iodide.
\newblock {\em Physical Review Letters}, 118, 10 2016.

\bibitem{Spiliotis_LSA}
Alexandros~K. Spiliotis, Michalis Xygkis, Michail~E. Koutrakis, Konstantinos
  Tazes, Gregoris~K. Boulogiannis, Chrysovalantis~S. Kannis, Georgios~E.
  Katsoprinakis, Dimitrios Sofikitis, and T.~Peter Rakitzis.
\newblock Ultrahigh-density spin-polarized hydrogen isotopes from the
  photodissociation of hydrogen halides: new applications for laser-ion
  acceleration, magnetometry, and polarized nuclear fusion.
\newblock {\em Light: Science {\&} Applications}, 10(1):35, Feb 2021.

\bibitem{Happer_1972}
William Happer.
\newblock Optical pumping.
\newblock {\em Rev. Mod. Phys.}, 44:169--249, Apr 1972.

\bibitem{Bouchiat_1969}
C.~C. Bouchiat, M.~A. Bouchiat, and L.~C.~L. Pottier.
\newblock Evidence for rb-rare-gas molecules from the relaxation of polarized
  rb atoms in a rare gas. theory.
\newblock {\em Phys. Rev.}, 181:144--165, May 1969.

\bibitem{Chlorine_HF}
Luther Davis, Bernard~T. Feld, Carrol~W. Zabel, and Jerrold~R. Zacharias.
\newblock The hyperfine structure and nuclear moments of the stable chlorine
  isotopes.
\newblock {\em Phys. Rev.}, 76:1076--1085, Oct 1949.

\bibitem{verdeyen}
J.T. Verdeyen.
\newblock {\em Laser Electronics}.
\newblock Laser electronics. Prentice Hall, 1995.

\bibitem{HCl_Cross_Section}
Bing-Ming Cheng, Chao-Yu Chung, Mohammed Bahou, Yuan-Pern Lee, and L.~C. Lee.
\newblock Quantitative spectral analysis of {HCl} and {DCl} in 120–220 nm:
  Effects of singlet–triplet mixing.
\newblock {\em The Journal of Chemical Physics}, 117(9):4293--4298, 2002.

\bibitem{Sofikitis_HClHBr_SPH}
Dimitris Sofikitis, Luis Rubio-Lago, Lykourgos Bougas, Andrew~J. Alexander, and
  T.~Peter Rakitzis.
\newblock Laser detection of spin-polarized hydrogen from {HCl} and {HBr}
  photodissociation: Comparison of h- and halogen-atom polarizations.
\newblock {\em The Journal of Chemical Physics}, 129(14):144302, 2008.

\end{thebibliography}
	\bibliographystyle{unsrt}
	
\end{document}